\documentclass[journal]{IEEEtran}
\IEEEoverridecommandlockouts
\usepackage{cite}
\usepackage{amsmath,amssymb,amsfonts}
\usepackage{textcomp}
\usepackage{url}
\usepackage{xcolor}
\usepackage{soul}
\usepackage{siunitx}
\usepackage{psfrag}
\usepackage{hyperref}       % hyperlinks
\usepackage{url}            % simple URL typesettings
\usepackage{booktabs}       % professional-quality tables
\usepackage{amsfonts}       % blackboard math symbols
\usepackage{nicefrac}       % compact symbols for 1/2, etc.
\usepackage{microtype}      % microtypography
\usepackage{xcolor}         % colors
\usepackage{soul}
\usepackage{algorithm}
\usepackage{algpseudocode}
\usepackage{floatpag}
\usepackage{enumitem}
\usepackage{arydshln}
\usepackage[hyphenbreaks]{breakurl}
\usepackage[pdftex]{graphicx}
\DeclareGraphicsExtensions{.pdf,.png,.jpg}
\usepackage{mathptmx}
\def\BibTeX{{\rm B\kern-.05em{\sc i\kern-.025em b}\kern-.08em
    T\kern-.1667em\lower.7ex\hbox{E}\kern-.125emX}}

\begin{document}

\title{
ReLaX-VQA: Residual Fragment and Layer Stack Extraction for Enhancing Video Quality Assessment
% Frame Differences Matter:
% Advanced Techniques for No-Reference Video Quality Assessment\\
% Frame Differences Matter in Quality Assessment of Compressed Videos
% ReLaX-VQA based on Residual Ensemble Feature Extraction 
\thanks{This work was funded by the UKRI MyWorld Strength in Places Programme (SIPF00006/1).}
}
\date{2024-08-01}

\author{\IEEEauthorblockN{Xinyi Wang,}
 \and
\IEEEauthorblockN{Angeliki Katsenou,}
\and
\IEEEauthorblockN{David Bull}

\IEEEauthorblockA{\textit{School of Computer Science, University of Bristol, Bristol, United Kingdom}} \\
\textbf{}
\{xinyi.wang, angeliki.katsenou, dave.bull\}@bristol.ac.uk}
\maketitle

\begin{abstract}
With the rapid growth of User-Generated Content (UGC) exchanged between users and sharing platforms, the need for video quality assessment in the wild is increasingly evident. UGC is typically acquired using consumer devices and undergoes multiple rounds of compression (transcoding) before reaching the end user. Therefore, traditional quality metrics that employ the original content as a reference are not suitable. In this paper, we propose ReLaX-VQA, a novel No-Reference Video Quality Assessment (NR-VQA) model that aims to address the challenges of evaluating the quality of diverse video content without reference to the original uncompressed videos. ReLaX-VQA uses frame differences to select spatio-temporal fragments intelligently together with different expressions of spatial features associated with the sampled frames. These are then used to better capture spatial and temporal variabilities in the quality of neighbouring frames.
Furthermore, the model enhances abstraction by employing layer-stacking techniques in deep neural network features from Residual Networks and Vision Transformers. Extensive testing across four UGC datasets demonstrates that ReLaX-VQA consistently outperforms existing NR-VQA methods, achieving an average SRCC of 0.8658 and PLCC of 0.8873. Open-source code and trained models that will facilitate further research and applications of NR-VQA can be found at: \url{https://github.com/xinyiW915/ReLaX-VQA}.
\end{abstract}


\section{Introduction}\label{intro}
% -Motivation\\
Video Quality Assessment (VQA) is a critical component when optimising user experience in video streaming applications. However, due to the vast volumes of User-Generated Content (UGC) on video-sharing platforms such as YouTube and TikTok~\cite{sandvine}, VQA is challenging. In the UGC use case, videos are typically first encoded on a user device (smartphone, consumer camera, etc.) when captured and then are transcoded by the streaming platform for storage and distribution.  Video parameters such as spatial resolution are often also modified to meet different user device or bandwidth constraints. As a result,  UGC video is more susceptible to degradation (via noticeable artefacts such as blocking, ringing, and blurring) than professional content~\cite{Adsumilli2019, wang2019youtube}. 

Traditional VQA methods that compare a distorted video with its pristine reference (e.g.) Peak Signal-to-Noise Ratio (PSNR) and Structural Similarity Index (SSIM)~\cite{wang2004image}) cannot be reliably used for UGC assessment, because the video uploaded to the sharing platform is not a pristine version. 
Similarly, learning-based full reference methods, such as Video Multi-Method Assessment Fusion (VMAF)~\cite{li2016toward}, cannot capture the quality of transcoded content because they have been designed and trained on pristine original references compared to content that has only been compressed once. 
As a consequence, there is an urgent need to develop robust No-Reference Video Quality Assessment (NR-VQA) models.

In recent years, Deep Neural Networks (DNNs) have been used to devise advanced NR-VQA models. The robust representation learning capability of DNNs has positioned them as the approach of choice for tackling a wide range of visual tasks~\cite{alzubaidi2021review}. Such models usually employ 2D-CNNs~\cite{li2019quality,tu2021rapique, ying2021patch}, 3D-CNNs~\cite{liu2018end, ying2021patch, sun2022deep}, and Transformers~\cite{wu2023discovqa,zhao2023zoom} and are designed to analyse pixel variations due to compression based on correlations with subjective quality scores. Existing models are primarily frame-based and struggle to capture temporal impairments effectively. Furthermore, the resizing of frames can result in further quality degradation, which has driven the development of VQA models that do not resize the frames but rather select frame fragments~\cite{wu2022fast, zhao2023zoom}. 
% Due to the scarcity of video datasets accompanied by subjective quality scores (mean opinion scores), some studies have explored strategies such as large cross-modal base models~\cite{wu2023exploring, wu2023towards, wu2023q}. 
Despite their efficiency, DNN VQA models are associated with high computational complexity and are data-hungry (requiring large volumes of data for training)~\cite{sze2017efficient}. 
Thus, the unavailability of large-scale datasets with subjective scores to train these networks from scratch limits their performance. 
% Ideally, the NR-VQA model used for UGC videos should exhibit a strong correlation with perceptual quality. 
% As mentioned above,  DNNs require a high amount of computational resources, high computational and memory requirements especially when processing high-resolution videos. This limits the testing efficiency and practical application~\cite{sze2017efficient}.
% 
Therefore, many existing VQA models rely on networks that are pre-trained on large image classification datasets to extract frame-based features~\cite{li2019quality,tu2021rapique,you2021long,madhusudana2023conviqt}. 
% This improves the efficiency and effectiveness of NR-VQA methods. 
Although the results look promising, this practice can result in a distributional shift between the original training task and the actual VQA application task. Moreover, despite the importance of frame-level spatial information, temporal information (motion or inter-frame variations) needs to be accounted for to evaluate the video quality accurately. 
% The spatio-temporal and motion information of the video cannot be effectively utilised if only relying on frame-level feature extraction. Instead, these features are important to improve the relevance of the VQA model to human perception.

In response to these challenges, we introduce ReLaX-VQA, a feature extractor that utilises two main techniques: (i) selective spatio-temporal fragmentation based on frame differences (or residuals) and optical flow and (ii) feature layer stacking. 
For the frame differences, inspiration is drawn from the structure of current hybrid block-based video encoders (e.g., H.264~\cite{r:h264} or H.265~\cite{r:HEVC}) where, for most coding modes frame encoding follows a hierarchical structure within a group of frames, with more bits allocated for the encoding of keyframes (I frames) and fewer for the successive predicted frames (P frames). Thus, the quality of the decoded frames varies per frame and is typically higher for I frames. Therefore, when subtracting a predicted from a key frame, the residual is an expression of both the temporal information in moving regions as well as spatial artefacts around stationary areas with compression artefacts. This residual information can be exploited to correlate it with the expected regions of visual attention that reflect perceptual variations in the Human Visual System (HVS). As this residual is sparse, we propose the selective extraction of fragments. The residual frames are split with a grid, and the selection is based on the ranking of residual fragments (using the absolute difference). This was designed on the basis that human attention is drawn to areas with visible distortions.
Additionally, fragmenting is also applied to temporal information, and particularly to the optical flow (OF) computed between the key and predicted frames. The OF can reveal regions within the frame with the most dynamic changes. This method can effectively utilise the spatio-temporal information to enhance the perception of the model. Based on the OF fragment selection, we also selected fragments from the corresponding locations in the sample video frame. 

The second technique relies on layer-stacking DNN feature extraction. Unlike deploying pooling techniques at a layer level that produces abstract representations, stacking features enhance the representation of deep features and increase their relevance to multi-layered human perception.
% The observation inspired this technique that, although most current models rely solely on feature extraction through average pooling in the final layer, as a result, it produces features that are abstract and difficult to interpret with human perceptual information. 
Through visualisation and perceptual correlation, we identified that different network layers capture information at different levels, from edges and textures to abstract features. This hierarchical feature extraction mechanism allows our model to mimic human visual processing more closely, thereby achieving more precise visual perception in complex scenarios. 
% improve visual processing.

Finally, to further improve the prediction performance of the extracted features, we constructed a simple and efficient Multi-Layer Perceptron (MLP) regression head for training and testing. By combining the characteristics of the Residual Network (ResNet)~\cite{he2016deep}, focusing on local information and the advantages of the Vision Transformer (ViT)~\cite{vaswani2017attention}, which concentrates on global features, our method improves the prediction accuracy. 

% We conducted experiments on large public video datasets to evaluate the performance of our model. By extracting residual fragments and stacking layers of deep features, we enhanced the prediction accuracy and correlation of our model. 
The primary contributions of this work are summarized below:
\begin{enumerate}[label=\roman*., itemsep=0.0em, leftmargin=*]
    \item \textbf{ReLaX-VQA Framework:} We proposed ReLaX-VQA, a novel NR-VQA model that employs two different strategies for feature extraction: selective fragmentation and layer stacking. Extensive evaluation of our model demonstrates higher accuracy compared to other state-of-the-art metrics while maintaining a relatively low computational complexity (see Section~\ref{sec: Experiments}).
    \item \textbf{Selective Feature Extraction:} We focus on selectively extracting ranked fragments from frame differences of key and predicted frames, from OF fields, and from the key frame. This biases the extraction of features from areas that are more prone to transcoding or scaling errors, thus enhancing spatio-temporal perception. Although the idea of fragmentation for NR-VQA has been used before~\cite{wu2022fast}, our implementation significantly differs through a dynamic selection based on ranked spatio-temporal differences between frames (further explained in Section~\ref{sec: bg}) and delivers improved results.
    \item \textbf{Incorporating Layer Stacking:} We boosted the feature abstraction capabilities by layer-stacked feature extraction from scaled (resized) frames using both ResNet and ViTs. The ResNets are better for a global feature representation, while ViTs have proven more effective for attention to salient regions. The layer stacking technique has proven effective in information fusion.
    % \item \textbf{Multi-Pooling:}
    \item \textbf{Improved MLP Regression:} We have improved the design of MLP regression that fuses local and global features for higher accuracy. We designed different MLP regressors for large-scale and small datasets for prediction. Moreover, we used an SGD optimiser and applied Cosine Annealing learning rate decay and Stochastic Weight Averaging to optimize the training process.
    
    % \hl{I would need a bit of information for this bullet. What was the main novelty you introduced compared to existing MLP regressors?}
    % \item We provide a visualisation of intermediate steps and analyse outliers to understand better the specific impacts of video features on VQA evaluation.
\end{enumerate}
The proposed model,  ReLaX-VQA, has been tested on several public large-scale datasets. The results, assessed by linear and rank correlation coefficients with mean opinion scores, Spearman (SRCC), Kendal (KRCC), and Pearson (PLCC), as well as Root Mean Square Error (RMSE), indicate higher accuracy on most datasets and consistently very high accuracy and correlation across datasets, outperforming existing NR-VQA methods. We have made the source code and pre-trained models available for public research and use.

The remainder of this paper is organized as follows. Section~\ref{sec: bg} summarises the related work in terms of existing datasets and NR-VQAs. Next, Section~\ref{sec: method} describes the proposed framework. Then, Section~\ref{sec: Experiments} details the experimental set-up, configurations, and results. Finally, Section~\ref{sec: Conlcusions} concludes this text with a summary of our findings and potential future directions.

\begin{figure*}[htbp]
    \centering
    \includegraphics[trim=0 50 0 102, clip, width=.9\textwidth]{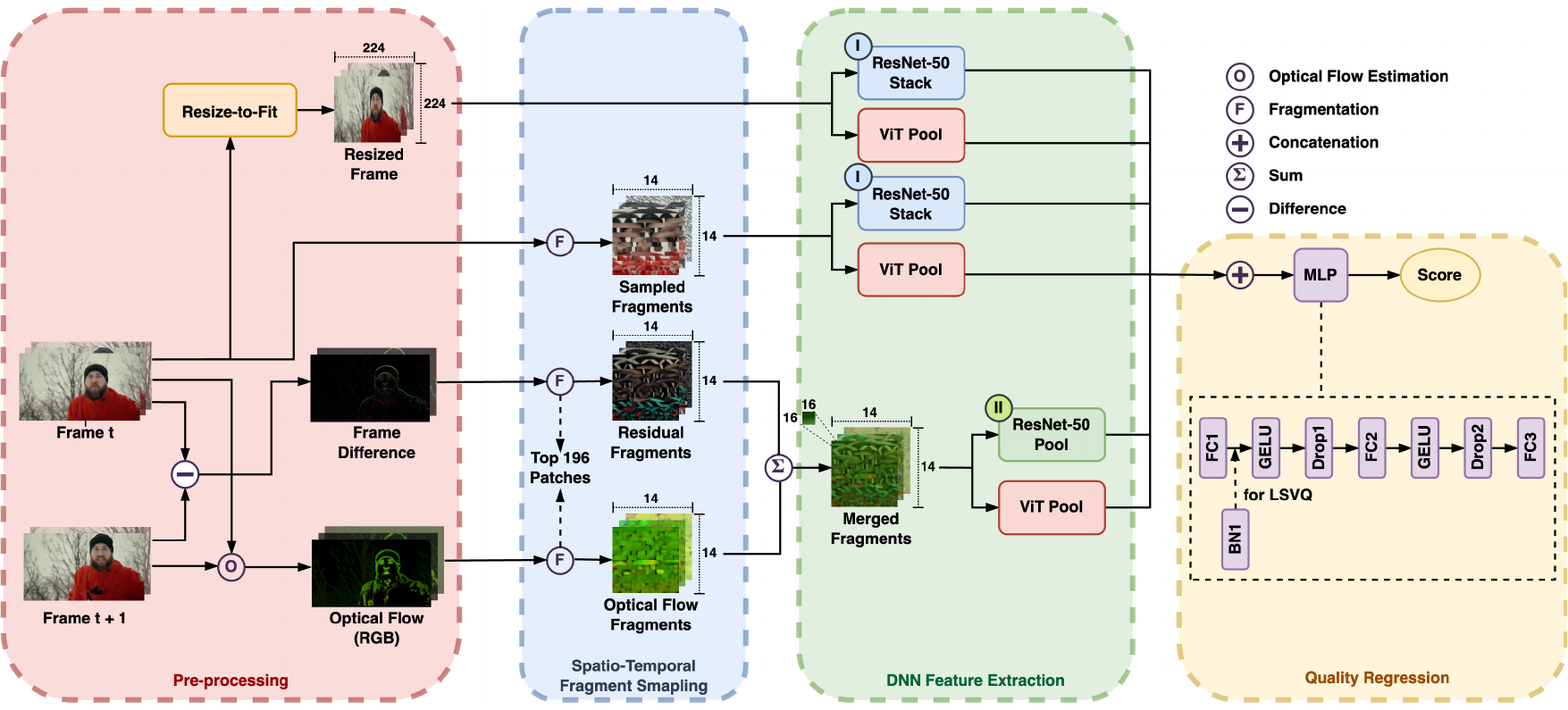}
      \vspace{-3.5em}
    \caption{Overview of the proposed ReLaX-VQA framework that demonstrates the different processing of the input frames for feature extraction and regression to infer the quality score. The visualisation of the intermediate representation of fragmented data is illustrated. More details on the architectures of ResNet-50 Stack (I) and ResNet-50 Pool (II) are provided in Fig. \ref{fig: framework_2}.}
  \label{fig: framework}
\end{figure*}

\section{Related work}
\label{sec: bg}

\begin{table}[t]
  \caption{Summary of VQA benchmark datasets}
  \label{tab:video-datasets}
  \centering
  \resizebox{\linewidth}{!}{%
  \begin{tabular}{@{}l|lllll@{}}
    \toprule
    Dataset & Resolution & Duration & Distortion & Subjective framework & Size\\
    \midrule
    CVD2014~\cite{nuutinen2016cvd2014} & 480p, 720p & 10-25s & In-capture & In-lab & 234\\
    % LIVE-Qualcomm~\cite{ghadiyaram2017capture} & 1080p & 15s & In-capture & In-lab & 208\\
    KoNViD-1k~\cite{hosu2017konstanz}& 540p & 8s & In-the-wild & Crowdsourced & 1,200\\
    LIVE-VQC~\cite{sinno2018large} & 240p-1080p & 10s & In-the-wild & Crowdsourced & 585\\
    YouTube-UGC~\cite{wang2019youtube} & 360p-4k & 20s & In-the-wild & Crowdsourced & 1,149\textsuperscript{*}\\
    LSVQ~\cite{ying2021patch} & 99p-4k & 5-12s & In-the-wild & Crowdsourced & 38,793\textsuperscript{**}\\
    \bottomrule
  \end{tabular}%
  }
\end{table}

\subsection{UGC video datasets}
To develop more realistic and accurate VQA models for UGC content, numerous datasets have been created. Table~\ref{tab:video-datasets} reports commonly used subjective video quality datasets used in recent studies. CVD2014~\cite{nuutinen2016cvd2014} and LIVE-VQC~\cite{sinno2018large} are datasets
% ~\cite{lin2015mcl, wang2016mcl, wang2017videoset} 
that simulate distorted videos from online video streaming through compression, transmission distortion, and artificial distortions (e.g., Gaussian blurring)~\cite{dodge2016understanding}. These datasets are small and contain only a limited number of unique source videos with restricted content diversity and distortion complexity. On the other hand, video quality databases collected ``in the wild'' specifically for UGC, are constantly being updated. These collections include a large number of real distorted videos produced by amateurs and more accurately reflect the complexities of UGC videos. Examples are the LIVE-Qualcomm~\cite{ghadiyaram2017capture}, KoNViD-1k~\cite{hosu2017konstanz}, YouTube-UGC~\cite{wang2019youtube}, and LIVE-VQC~\cite{sinno2018large}. Although these datasets are larger and contain more content variation, they are still insufficient to train deep learning models. The recent release of two large-scale VQA datasets, FlickrVid-150k~\cite{gotz2019no} (not publicly available) and LSVQ~\cite{ying2021patch} tried to address the lack of training data for deep learning-based VQA models. Notably, the publicly available LSVQ dataset, which contains 39,076 videos, demonstrates advantages in handling large-scale video data. 
% LSVQ also provides two official test subsets: LSVQ\(_{test}\) and LSVQLSVQ\(_{test_1080p}\).

\subsection{NR VQA models}
% \paragraph{Handcrafted feature-based VQA}
Numerous classic NR-VQA models have been developed based on traditional Natural Scene Statistics (NSS), such as NIQE~\cite{mittal2012making}, PIQE~\cite{moorthy2011blind}, and BRISQUE~\cite{mittal2011blind}.
% focus respectively on the statistical features of images, simulating the perceptual quality of the HVS, and spatial domain NSS. 
Inspired by frequency domain characteristics, V-BLIINDS~\cite{6705673} and V-CORNIA~\cite{7025098} used Discrete Cosine Transform (DCT) statistics and learned codebooks to mark image integrity and block features, respectively. These methods perform better for specific types of distortions (e.g., blurriness) and have been designed to capture a specific range of video quality. Other models employ a large number of handcrafted features and feature selection strategies to evaluate video quality. Successful examples are TLVQM~\cite{korhonen2019two} and VIDEVAL~\cite{tu2021ugc}. TLVQM combines high spatial complexity and low temporal complexity features, while VIDEVAL integrates a wide range of handcrafted features to simulate real distortions. These models train shallow regression models with statistical features and perform well on standard video (not UGC) content but are limited in generalising over UGC content.

% \paragraph{Learning-based VQA} 
More recently, VQA research has shifted towards DNNs, employing CNNs and Transformers for constructing feature extractors. Notably, RAPIQUE~\cite{tu2021rapique} combines NSS with semantic-aware CNN features extracted from downsampled frames, resulting in reduced computational runtime.
% VQA encompasses temporal complexity and requires advanced networks to process spatial and temporal dynamics. 
Other high-performing models include VSFA~\cite{li2019quality} and its enhanced version, MDTVSFA~\cite{li2021unified} (using a hybrid dataset training strategy). These extract frame features using a pre-trained image classification model ResNet-50~\cite{he2016deep} and apply gated recursive units to process temporal features. The latest developments in NR-VQA focus on addressing multi-resolution and complex spatio-temporal distortions in UGC. Using patches and fragments as in~\cite{ying2021patch, wu2022fast, zhao2023zoom} and/or novel data sampling mechanisms as in~\cite{liu2024scaling, ke2023mret} helps preserve the global and local details of high-resolution videos, capturing complex spatio-temporal features and improving processing efficiency. Other NR-VQA models perform end-to-end feature training or multi-scale feature fusion, utilising attention mechanisms and simulating HVS features to enhance performance~\cite{wu2022fast, lao2022attentions, zhang2022hvs, sun2022deep}. A powerful aspect of VQA metrics is the ability to rank quality levels. Chen et al~\cite{Feng_2024_WACV} focus on improving the ranking performance of their model by using VMAF-based quality ranking information but train their model on standard video datasets with access to uncompressed content. Li et al.~\cite{li2022blindly} address domain bias by transferring knowledge from image quality and motion recognition databases to improve benchmarks. UVQ~\cite{wang2021rich} further augments the video quality metrics by incorporating semantic information. 
% % add on：
% Traditional VQA models extensively use classic machine learning methods such as Support Vector Regression (SVR) to map visual features to quality scores, but these require substantial annotated data and computational resources. 
% DNNs provide improved accuracy and adaptability, eliminating the need to handcraft features and efficiently processing distorted images in NR image quality assessment (IQA)~\cite{}. RAPIQUE integrates NSS with deep features from CNNs, improving performance on large UGC datasets while reducing computational requirements.  

\subsection{Advancing the State-of-the-Art}
Two of the best-performing learning-based approaches mentioned above that were designed for the quality assessment of UGC content are VSFA~\cite {li2019quality} and Fast-VQA~\cite{wu2022fast, wu2023neighbourhood}. These will comprise the benchmarks of the learning based methods for our experiments.
VSFA~\cite{li2019quality} utilises content-aware features of pre-trained neural networks and gated recursive units to model temporal dependencies. However, the fixed feature extraction method may not fully capture the diverse quality variations of UGC content. In contrast, ReLaX-VQA embodies dynamic fragment selection based on frame differences to more effectively capture temporal variations. This enhances robustness and accuracy through deep learning feature extraction. ReLaX-VQA achieves higher linear and rank correlation, aligning more closely with human perception.

Wu et al. proposed Fast-VQA and DOVER models\cite{wu2022fast, wu2023neighbourhood, wu2023dover} that apply fragmentation using a uniform grid to randomly select fragments with the benefit of reduced computational costs. However, the uniform sampling strategy for fragmentation may omit perceptually important spatio-temporal changes in UGCs. Compared to Fast-VQA, DOVER provides the option of multimodality through an aesthetics perspective. ReLaX-VQA further advances this technique by employing a ranking of the spatio-temporal fragments based on pixel differences between key and predicted frames, using mean absolute differences to capture significant quality variations. As a result, ReLaX-VQA enhances both temporal and spatial perception, demonstrating higher accuracy and better alignment with human perception compared to existing NR-VQA methods. The results reported in Section~\ref{ssec: Performance} confirm the effectiveness of our proposed method.

% \hl{What are the main differences of the proposed ReLaX-VQA when compared to state-of-the-art? Provide a small paragraph explaining what is similar and what you do differently. Do that for Fast-VQA and whichever learning-based metric you think is more appropriate, e.g. VSFA
% Example to start with:\\
% The idea of fragmentation for NR-VQA has been proposed before by Wu et al~\cite{wu2022fast}. Their implementation is based on a random selection of patches within a grid... We propose an advanced technique that uses the pixel differences between key and predicted frames to rank the fragments based on the mean absolute differences.

\section{Proposed method}
\label{sec: method}
\begin{figure*}[htbp]
  \centering
  \includegraphics[width=\linewidth]{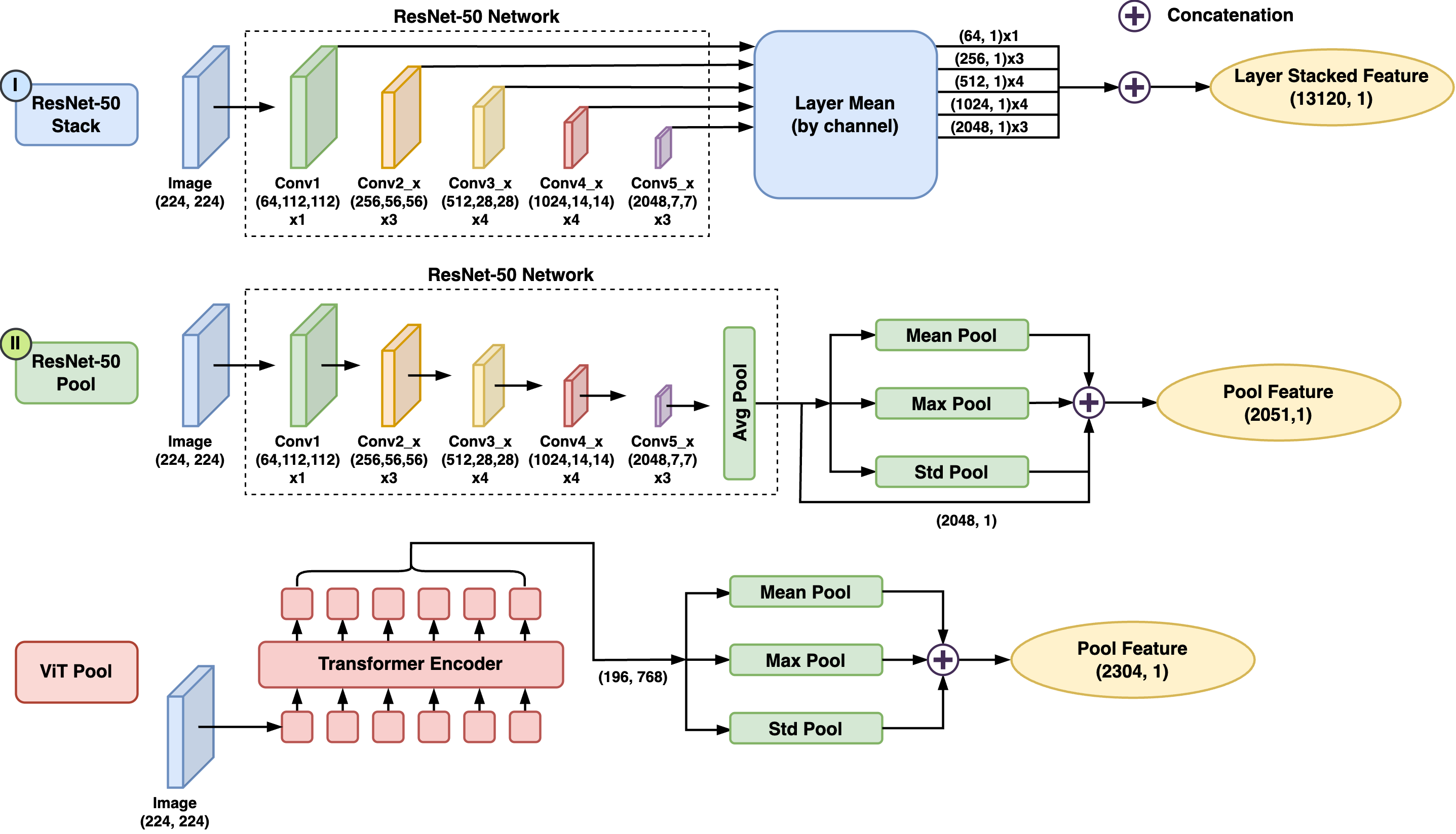}
  \caption{The bespoke architectures of ResNet-50 Stack (I),  ResNet-50 Pool (II), and ViT Pool.}
  \label{fig: framework_2}
\end{figure*}

An overview of the proposed ReLaX-VQA model framework is provided in Fig.~\ref{fig: framework} and consists of three basic modules: the Spatio-Temporal Fragment Sampling module (Section~\ref{ssec: residual}), the DNN Feature Extraction module (Section~\ref{ssec: stack}), and Quality Regression module (Section~\ref{ssec: regression}). Firstly, we extract features from successively sampled video frames using the spatio-temporal fragment sampling module: frames, frame differences (residuals), and optical flow. We also, resize the frames and extract DNN features from the entire frame. Then, we stack the extracted features from the resized and fragmented frame using different layers of the ResNet-50, thus fusing these multi-layered features to capture the quality-aware features. Here, we note that since the ViT is based on a self-attention mechanism, we do not use a layer-stacking framework for this, as also shown in Fig.~\ref{fig: framework_2}. Finally, a simple MLP quality regression module maps the quality-aware features to video quality scores. Furthermore, we introduced a hybrid Mean Absolute Error (MAE) and Rank loss function to optimise the quality prediction model. Through this approach, ReLaX-VQA can effectively utilise the residual information after dynamic changes between video frames. The local and global information that affects the video quality is captured through feature layer stacking, enhancing the model's abstraction capability and its consistency with human perception.

\subsection{Spatio-Temporal Fragment Sampling module}
\label{ssec: residual}
Frequently in VQA, researchers resize or crop frames prior to feature extraction. Resizing often introduces quality degradation as it filters or changes local textures, which are often critical in the perception of video quality, while cropping affects the alignment of global and local quality. Particularly at high spatial resolutions, these processing methods might result in a significant loss of detail. Moreover, the viewer does not only notice the overall quality but also temporal artefacts, e.g., sudden object transitions/deformations due to motion estimation errors, and/or any other spatial artefacts, e.g., blocking artefacts, colour bleeding. By extracting the key fragments with the highest dynamic changes from frame to frame, we can effectively capture both local and global information affecting video quality. Most UGC videos are short, usually containing only one shot and redundancy in spatio-temporal information, especially at high frame rates. Therefore, we propose an innovative Residual Fragment (RFs) sampling method to capture the effect of temporal and spatial variations on video quality. 

% % frame difference
We sampled the video at a rate of two frames every half-second to obtain successive and temporally correlated frame pairs \(F_{t}\) and \(F_{t+1}\). We first computed the frame difference \(\Delta F_t(x, y)\) by calculating the absolute difference of each pixel in between the extracted frame pairs to capture the inter-frame content variations, as follows:
\begin{equation}
    \Delta F_t(x, y) = |F_{t+1}(x, y) - F_t(x, y)| ,
\label{eq:fd}
\end{equation} 
where \((x, y)\) represents the pixel position, and \(|\cdot|\) denotes the absolute value. In most scenes, there is less variation in the static background between successive frames, whereas moving objects or shifts in the background caused by camera movement can produce higher pixel differences.

% % optical flow
Concurrently, to better capture motion information, we applied Farneback dense optical flow estimation~\cite{farneback2003two} to quantify pixel displacement across video frames based on temporal continuity and spatial luminance consistency. Optical flow \(f_t(x,y)\) is computed by comparing the pixel positions between consecutive frames \(F_{t}\) and \(F_{t+1}\) to estimate the direction and velocity of movement of each reference pixel \((x,y)\), by solving the following equation:
\begin{equation}
    F_t(x, y) \approx F_{t+1}(x + \delta x, y + \delta y) ,
\label{eq:of}
\end{equation}
where \(\mathbf{\delta}x\) and \(\mathbf{\delta}y\) are the horizontal and vertical components, representing the predicted movement distance of pixel \((x,y)\). For easier visualisation of the motion in the colour space, we converted the magnitude and angle of the optical flow vectors into colour information and transformed the colour space from HSV to RGB:
\begin{multline}
    \text{\small RGB} = \text{\small HSVtoRGB}\left(\left[\frac{\arctan2(v, u)}{\pi} \times 180, \, 2^b-1, \right.\right. \\
    \left.\left.\min\left(\frac{\sqrt{u^2 + v^2}}{\max(|\mathbf{F}|)}, 1\right) \times (2^b-1)\right]\right) ,
\label{eq:of2}
\end{multline}
Where \(\mathbf{F}\) represents the entire optical flow field, with the optical flow vector at each point consisting of horizontal components \(u\) and vertical components  \(v\), and \(b\) is the bit-depth.

% % top patches and RFs
\begin{algorithm}[t] 
\caption{Spatio-temporal Residual Fragment Extraction}
\label{alg: rf}
\begin{algorithmic}[1]

\State \textbf{Input:} Frame difference or optical flow \(F\), Patch size \(p\), Target image size \(n\), Top-\(K\) patches
\State \textbf{Output:} Patches matrix \(P\), Patch positions \(P_\text{positions}\), Residual fragments \(RF\), 

\State Initialise matrix \(P \gets 0\)
\For{$i = 0$ to $height - p$}
    \For{$j = 0$ to $width - p$}
        \State Extract patch: \( \text{patch} \gets F[i:i+p, j:j+p] \)
        \State Calculate sum: 
        \State \quad\quad\quad\quad \( P_{\Delta F_t}(i,j) \gets  \eqref{eq:sf1} \) \Comment{For frame difference}
        \State \quad\quad\quad\quad \( P_{f_t}(i,j) \gets \eqref{eq:sf2} \) \Comment{For optical flow}
    \EndFor
\EndFor

\State Find top-\(K\) patches: \(top\_idx \gets \text{sort}(\text{argsort}(P, K))\)

\State Initialise \(RF \gets 0\) and \(P_\text{positions} \gets []\)
\For{each index \((y, x)\) in \(top\_idx\)}
    \State Extract patch: 
    \State \(patch \gets P[y \cdot p:(y+1) \cdot p, x \cdot p:(x+1) \cdot p]\)
    \State Determine new location in the target image \(I_\text{patches}\) (size: \(n \times n\)):
    \[
    \text{row\_idx} \gets \left\lfloor \frac{\text{idx}}{\left\lfloor \frac{n}{p} \right\rfloor} \right\rfloor, \quad \text{col\_idx} \gets \text{idx} \bmod \left\lfloor \frac{n}{p} \right\rfloor
    \]
    \State Insert patch into \(RF\) 
    \State and update \(P_\text{positions} \gets P_\text{positions} \cup \{(y, x)\}\)
\EndFor

\State \Return \(P, RF, P_\text{positions}\)

\end{algorithmic}
\end{algorithm}

% % top patches
After computing \(\Delta F_{t}\) and \(f_t(x,y)\), we extracted regions with high residual information, termed ``top patches". We uniformly segmented the frame difference and optical flow into fixed-size pixel patches \((p \times p)\). For each patch, we computed the following sums:
\begin{equation}
\label{eq:sf1}
    S_{\Delta F_t}(i,j) = \sum_{x=i}^{i+p} \sum_{y=j}^{j+p} \Delta F_t(x, y) , 
\end{equation}
\begin{equation}
\label{eq:sf2}
    S_{f_t}(i,j) = \sum_{x=i}^{i+p} \sum_{y=j}^{j+p} \|\mathbf{f}_t(x, y)\| ,
\end{equation}
where \(S_{\Delta F_t}(i,j) \) and \(S_{f_t}(i,j)\) are the sum of frame differences and the sum of optical flow magnitudes \( \|\mathbf{f}_t(x, y)\|\) in the patch \((i,j)\), respectively.
% % Residual fragment Extraction

The selected top patches from the frame differencing and optical flow are mapped into a merged fragment, MF, which retains the most salient regions of variation in the time series while discarding the static or low residual information:
\begin{equation}
\label{eq:mf}
    MF(x, y) = \alpha \cdot RF_{\Delta F}(x, y) + (1 - \alpha) \cdot RF_{f}(x, y) ,
\end{equation}
where \(MF(x, y)\) is the merged fragment and \(\alpha=0.5\) is a weighting parameter for the two fragments, \(RF_{\Delta F}\) and \(RF_{f}\). 
The target RFs are sized at \(224 \times 224\) pixels to align with the model input requirements. We set the size of the top patches to \( p \times p \) (where \( p = 16 \)) pixels, arranging them in a \(14\times14\) grid to achieve the target dimensions. Consequently, a total of 196 important patches need to be selected. This process is detailed in Algorithm~\ref{alg: rf}(line 11). We also extracted patches from the same positions as the top frame difference patches to construct sampled spatial fragments based on the principle of spatial position invariance. This sampling criterion is based on the hypothesis that patches with higher residual information (indicating increased movement or variation) are critical for understanding changes in video content and are more likely to convey noticeable compression artefacts. 

\begin{figure}[htbp]
  \centering
  \includegraphics[width=\linewidth]{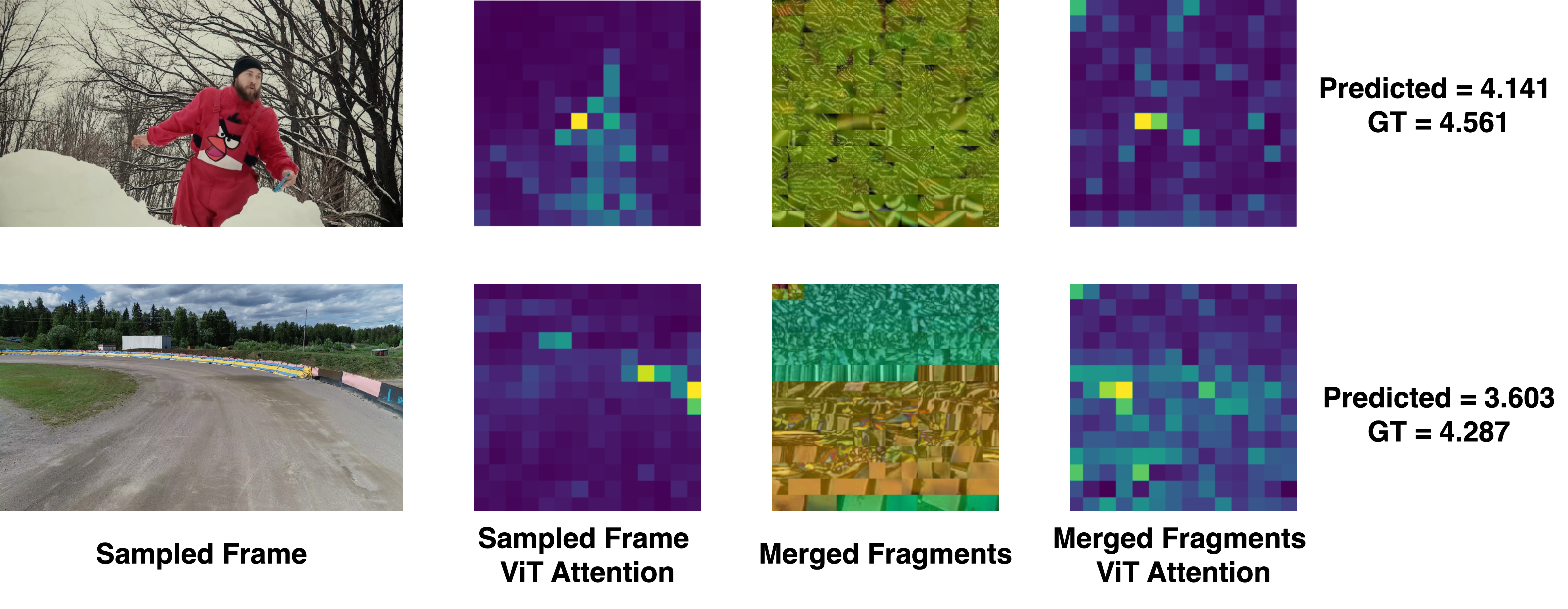}
  \caption{Illustration of examples sampled from the YouTube-UGC~\cite{wang2019youtube} dataset. \textbf{Top-row}: \(TelevisionClip\_1080P-68c6.mkv\). 
  \textbf{Bottom-row}: \(Sports\_2160P-0455.mkv\). GT refers to ground truth quality scores. Shown here are the scores predicted using \textbf{ReLaX-VQA (w/o FT).}}
  \label{fig: FragmVisualisation}
\end{figure}

We also provide frames from two example videos to visualise the merged fragments in Fig.~\ref{fig: FragmVisualisation}.
Here, we present the sampled frames and their corresponding ViT attention visualisations, along with the merged fragments (ours) and their ViT attention. These visualisations highlight the attention focus of the model. As shown in Fig.~\ref{fig: FragmVisualisation}, the first video, \(TelevisionClip\_1080P-68c6.mkv\), is a clip showing a man wearing an Angry Birds print moving in a snowy environment. The second video \(Sports\_2160P-0455.mkv\) is a scene from a camera positioned slightly above the dashboard of a racing car on a track, providing a clear view through the windshield. The scene shows changes in perspective, including significant jittering. The safety barriers on the curve appear moving faster. The videos can be found in the YouTube-UGC dataset~\cite{wang2019youtube}.
% at \href{https://media.withyoutube.com/}{https://media.withyoutube.com/}. 
It can be observed that we get good results even when using our frozen ReLaX-VQA model without fine-tuning (ReLaX-VQA (w/o FT) to predict video quality scores.

\subsection{DNN Feature Extraction module}
\label{ssec: stack}
We observed that the lower (first) layers of CNN networks process basic image elements such as edges, colours, and textures, while deeper layers handle more advanced abstract information. Recent studies~\cite{wang2023ugc, ramsookicip2023} indicated that saliency features may be inherent in representations of different DNN layers (not in just pooled flattened features), allowing us to simulate human attention to regions of interest. Therefore, in this work, we used layer stacking and global pooling techniques to enhance DNN features and improve the perceptual accuracy of the prediction model. Besides this, through the integration of ViT features~\cite{vaswani2017attention}, which focus on global information, with ResNet-50~\cite{he2016deep} features, which focus on local details, we can further improve the model's ability to process complex visual information. To support this, we conducted an ablation study to evaluate the impact of using different features alone, further validating the proposed method (see Section~\ref{sec: Experiments}, Table~\ref{tab: AblationStackVSPool}).

% % Stack
For each layer \(n\) of ResNet-50, the feature map \(m^{(n)}(x)\) has dimensions \(C_n \times H_n \times W_n\), where \(C_n\) represents the number of channels, \(H_n\) represents the height, and \(W_n\) represents the width. The channel mean is computed for each channel \(c\) in the \(n\) layer. The feature vectors \(v^{(C_n)}\) from each layer are concatenated to form a layer-stacked feature comprising all the layer features:
\begin{equation}
    v^{(n)}c = \frac{1}{H_n \times W_n} \sum{h=1}^{H_n} \sum_{w=1}^{W_n} m^{(n)}(x_{c,h,w}) ,
\end{equation}
\begin{equation}
    V_{\text{CNN}}^{\text{stack}} = v^{(1)} \oplus v^{(2)} \oplus \cdots \oplus v^{(C_l)} .
\end{equation}
% % Pool
Along with layer stacking, we also applied global pooling components to the network, computing average pooling, maximum pooling and standard pooling:
\begin{equation}
    \mu = \text{mean}(m^{(n)}(x)), \zeta = \text{max}(m^{(n)}(x)), \sigma = \text{std}(m^{(n)}(x)) .
\end{equation}
Here, \(m(x)\) is defined differently in different networks. In ResNet-50, \(m(x)\) refers to the feature \((\text{avg\_pool})\) following the average pooling layer of the network. Consequently, the enhanced features are shaped by computing global statistics and concatenating them with the \(\text{avg\_pool}\) feature vectors. In contrast, in ViT, \(m(x)\) refers to the patch embeddings (excluding CLS tokens) output by the encoder, with dimensions \(N \times D\), where \(N\) denotes the number of patches and \(D\) the dimension of each patch. We performed global pooling directly on the patch embeddings:
\begin{equation}
    V_{\text{CNN}}^{\text{pool}} = V_{\text{avg\_pool}} \oplus \mu_{\text{avg\_pool}} \oplus \zeta_{\text{avg\_pool}} \oplus \sigma_{\text{avg\_pool}}
\end{equation}
\begin{equation}
    V_{\text{ViT}}^{\text{pool}} = \mu_{\text{patch\_emb}} \oplus \zeta_{\text{patch\_emb}} \oplus \sigma_{\text{patch\_emb}}
\end{equation}
% % Combined
Next, we aggregated the enhanced feature vectors from ResNet-50 and ViT to acquire DNN features for the sampled video frames. After processing all the sampled frames, we derived a feature representation of the video sequence by averaging the features of each frame:
\begin{equation}
    V_{\text{video}} = \frac{1}{N} \sum_{i=1}^N (V_{\text{CNN}} \oplus V_{\text{ViT}}).
\end{equation}

\subsection{Quality Regression module}
\label{ssec: regression}
We designed a simple and efficient MLP~\cite{rosenblatt1958perceptron} regression head to predict video quality scores. The model consists of three fully connected layers, with hidden layers of dimensions 256 and 128, and the last layer outputs the predicted scores. We applied the GELU activation function and a dropout rate of 0.1 after the first two layers, with the final layer directly outputting the results. Throughout the training process, we utilized the SGD optimizer and applied Cosine Annealing learning rate decay~\cite{loshchilov2016sgdr}. Additionally, we introduced the Stochastic Weight Averaging (SWA)~\cite{izmailov2018averaging} technique to enhance the model's generalisation performance. To prevent overfitting, we adopted an early stopping strategy with a patience of 5 epochs. Moreover, we incorporated batch normalization for handling the large dataset LSVQ~\cite{ying2021patch}.

\subsection{Loss function}
\label{ssec: loss}
We used a composite loss function inspired by ~\cite{wen2021strong}, where the MAE loss and Rank Loss are weighed by specific weights to achieve optimal learning. Specifically, we introduced two weighting parameters, \(l1_w\) and \(rank_w\), to adjust the weights of the two losses in the total loss function. MAE loss is used to calculate the average absolute difference between predicted scores and ground truth:
\begin{equation}
    L_{\text{MAE}} = \frac{1}{n} \sum_{i=1}^n |y_{\text{pred}, i} - y_{\text{true}, i}| ,
\end{equation}
where \(y_i \in \mathbb{R}\) represents a quality score, \(n\) is the number of videos.

Rank loss encourages the model to maintain the correct ranking by comparing the differences between paired predicted and true values. Since scores on the diagonal compare to themselves, here we excluded the effect of rank on the diagonal. The rank loss \(L_{\text{Rank}}\) can be expressed as:
\begin{equation}
    L_{\text{Rank}}^{\text{ij}} = \max\left(0, \delta_{ij} - e(y_{\text{true}, i}, y_{\text{true}, j}) \cdot d_{ij}\right) ,
\end{equation}
\begin{equation}
    L_{\text{Rank}} = \frac{1}{n \cdot (n-1)} \sum_{i=1}^n \sum_{j=1}^n L_{\text{Rank}}^{\text{ij}} ,
\end{equation}
where \(\delta_{ij} = |y_{\text{true}, i} - y_{\text{true}, j}|\) represents the absolute difference in true scores, and \(d_{ij} = y_{\text{pred}, i} - y_{\text{pred}, j}\) represents the difference in predicted scores. The function \(e(y_{\text{true}, i}, y_{\text{true}, j})\) is defined as follows:

\begin{equation}
    e(y_{\text{true}, i}, y_{\text{true}, j}) =
\begin{cases}
\,\,\,\, 1 & \text{, if } y_{\text{true}, i} \geq y_{\text{true}, j} \\
-1 & \text{, otherwise.}
\end{cases}
\end{equation}

Finally, the total loss \(L_{\text{Total}}\) helps the model to capture more information about different distortions or different details, thus speeding up  convergence, calculated as:
\begin{equation}
    L_{\text{Total}} = L_{\text{MAE}} \times l1_w + L_{\text{Rank}} \times rank_w \, .
\end{equation}

\begin{table*}[t]
\centering
\caption{Performance comparison of the evaluated NR-VQA models on the four NR-VQA datasets. The \textcolor{red}{\textbf{Red}}, \textcolor{blue}{\textbf{blue}}, and \textbf{boldface} entries indicate the 1st, 2nd and 3rd best performance on each database for each performance metric, respectively. The $\dagger$ denotes the exclusion of 4K videos.}
\label{tab: ComparisonToSoA}
\begin{tabular}{@{}lcccccccccccccccc@{}}
\toprule
\textbf{Datasets}  & \multicolumn{4}{c}{\textbf{CVD2014~\cite{nuutinen2016cvd2014}}} & \multicolumn{4}{c}{\textbf{KoNViD-1k~\cite{hosu2017konstanz}}}\\
\cmidrule(lr){2-5} \cmidrule(lr){6-9}
\textbf{Models/Metrics} & \textbf{SRCC} & \textbf{KRCC} & \textbf{PLCC} & \textbf{RMSE} & \textbf{SRCC} & \textbf{KRCC} & \textbf{PLCC} & \textbf{RMSE}\\
\midrule
BRISQUE~\cite{mittal2011blind} & 0.5553 & 0.3895 & 0.5527 & 18.4752 & 0.6781 & 0.4935 & 0.6746 & 0.4753 \\
V-BLIINDS~\cite{6705673} & 0.7306 & 0.5393 & 0.7853 & 13.7267 & 0.7258 & 0.5322 & 0.7155 & 0.4498 \\
TLVQM~\cite{korhonen2019two} & 0.5399 & 0.4006 & 0.5785 & 18.0832 & 0.7616 & 0.5635 & 0.7463 & 0.4285 \\
VIDEVAL~\cite{tu2021ugc} & 0.7663 & 0.5634 & 0.8062 & 13.1151 & 0.8073 & 0.6036 & 0.7923 & 0.3928 \\
\hdashline
\noalign{\vskip 3pt}
RAPIQUE~\cite{tu2021rapique} & 0.8530 & 0.6836 & 0.8766 & 10.6670 & 0.8219 & 0.6264 & 0.8191 & 0.3693 \\
VSFA\(\dagger\)~\cite{li2019quality} & \textcolor{blue}{\textbf{0.8825}} & \textcolor{blue}{\textbf{0.7179}}& 0.8784 & \textbf{9.8619} & 0.8067 & 0.6102 & 0.8182 & 0.4056 \\
Fast-VQA-B (w/o FT)~\cite{wu2022fast} & 0.8124 & 0.6125 & 0.8221 & 12.7920 & 0.8581 & 0.6684 &  0.8551 & 0.3450 \\
Fast-VQA-B (w/ FT)~\cite{wu2022fast} & 0.8341 & 0.6466 & 0.8456 & 11.6044 & \textcolor{red}{\textbf{0.8869}} & \textcolor{red}{\textbf{0.7052}} & \textcolor{blue}{\textbf{0.8776}} & \textcolor{blue}{\textbf{0.3131}} \\
DOVER (Technical Branch)~\cite{wu2023dover} & 0.8571 & \textbf{0.6966} & \textbf{0.8588} & 11.0493 & \textcolor{blue}{\textbf{0.8736}} & \textcolor{blue}{\textbf{0.6962}} & \textcolor{red}{\textbf{0.8852}} & \textcolor{red}{\textbf{0.2920}} \\
\midrule
\textbf{ReLaX-VQA (ours)} & \textbf{0.8643} & 0.6960 & \textcolor{blue}{\textbf{0.8895}} & \textcolor{blue}{\textbf{9.8185}} & 0.8535 & 0.6594 & 0.8473 & 0.3370 \\
\textbf{ReLaX-VQA (w/o FT)} & 0.7845 & 0.5930 & 0.8336 & 12.2445 & 0.8312 & 0.6418 & 0.8427 & 0.3466 \\
\textbf{ReLaX-VQA (w/ FT)} & \textcolor{red}{\textbf{0.8974}} & \textcolor{red}{\textbf{0.7299}} & \textcolor{red}{\textbf{0.9294}} & \textcolor{red}{\textbf{8.1812}} & \textbf{0.8720} & \textbf{0.6881} & \textbf{0.8668} & \textbf{0.3211} \\
\bottomrule
\end{tabular}

\begin{tabular}{@{}lcccccccccccccccc@{}}
\toprule
\textbf{Datasets}  & \multicolumn{4}{c}{\textbf{LIVE-VQC~\cite{sinno2018large}}} & \multicolumn{4}{c}{\textbf{YouTube-UGC~\cite{wang2019youtube}}}\\
\cmidrule(lr){2-5} \cmidrule(lr){6-9}
\textbf{Models/Metrics}& \textbf{SRCC} & \textbf{KRCC} & \textbf{PLCC} & \textbf{RMSE} & \textbf{SRCC} & \textbf{KRCC} & \textbf{PLCC} & \textbf{RMSE}\\
\midrule
BRISQUE~\cite{mittal2011blind} & 0.6096 & 0.4420 & 0.6652 & 12.7480 & 0.3517 & 0.2416 & 0.3768 & 0.6349 \\
V-BLIINDS~\cite{6705673} & 0.6925 & 0.5046 & 0.6872 & 12.4031 & 0.4783 & 0.3396 & 0.4967 & 0.5949 \\
TLVQM~\cite{korhonen2019two} & 0.8133 & 0.6231 & 0.7912 & 10.4409 & 0.6802 & 0.4892 & 0.6876 & 0.4977 \\
VIDEVAL~\cite{tu2021ugc} & 0.7725 & 0.5874 & 0.7752 & 10.7846 & 0.7814 & 0.5906 & 0.7929 & 0.4177 \\
\hdashline
\noalign{\vskip 3pt}
RAPIQUE~\cite{tu2021rapique} & 0.7328 & 0.5418 & 0.7520 & 11.2538 & 0.7994 & 0.5989 & 0.8300 & 0.3823 \\
VSFA\(\dagger\)~\cite{li2019quality} & 0.5898& 0.4120& 0.5894& 15.3245 & 0.7857& 0.5814& 0.7808& 0.4321 \\
Fast-VQA-B (w/o FT)~\cite{wu2022fast} & \textbf{0.8261} & \textbf{0.6344} & 0.8466 & 9.4487 & 0.7241 & 0.5266 & 0.7364 & 0.4728 \\
Fast-VQA-B (w/ FT)~\cite{wu2022fast} & \textcolor{red}{\textbf{0.8484}} & \textcolor{red}{\textbf{0.6658}} & \textcolor{blue}{\textbf{0.8703}} & \textcolor{blue}{\textbf{8.8464}} & \textcolor{blue}{\textbf{0.8385}} & \textcolor{blue}{\textbf{0.6444}} & \textbf{0.8302} & 0.3880 \\
DOVER (Technical Branch)~\cite{wu2023dover} & 0.7898 & 0.6098 & \textbf{0.8573} & \textbf{9.0036} & \textbf{0.8181} & \textbf{0.6290} & 0.8197 & \textcolor{blue}{\textbf{0.3559}} \\
\midrule
\textbf{ReLaX-VQA (ours)} & 0.7655 & 0.5785 & 0.8079 & 9.8596 & 0.8014 & 0.6167 & 0.8204 & 0.3801 \\
\textbf{ReLaX-VQA (w/o FT)} & 0.7664 & 0.5812 & 0.8242 & 9.8201 & 0.8104 & 0.6131 & \textcolor{blue}{\textbf{0.8354}} & \textbf{0.3768} \\
\textbf{ReLaX-VQA (w/ FT)} & \textcolor{blue}{\textbf{0.8468}} & \textcolor{blue}{\textbf{0.6649}} & \textcolor{red}{\textbf{0.8876}} & \textcolor{red}{\textbf{7.9869}} & \textcolor{red}{\textbf{0.8469}} & \textcolor{red}{\textbf{0.6623}} & \textcolor{red}{\textbf{0.8652}} & \textcolor{red}{\textbf{0.3437}} \\
\bottomrule
\end{tabular}
\end{table*}

\section{Experiments}
\label{sec: Experiments}

\subsection{Experimental setup}
\label{ssec: ExpSetup}
\paragraph{UGC Datasets} Our model's performance has been evaluated on current state-of-the-art NR-VQA datasets including CVD-2014~\cite{nuutinen2016cvd2014}, KoNViD-1k~\cite{hosu2017konstanz}, LIVE-VQC~\cite{sinno2018large}, and YouTube-UGC~\cite{wang2019youtube}. Additionally, we have conducted feature extraction and training on the large-scale LSVQ~\cite{ying2021patch} dataset and successfully migrated to smaller VQA datasets by Fine-Tuning (FT) the pre-trained model, thereby achieving better performance. We excluded 46 greyscale videos from the YouTube-UGC\textsuperscript{*} dataset, leaving 1,103 videos, and 283 greyscale videos from the LSVQ\textsuperscript{**} dataset, leaving 38,510 videos.

\paragraph{Evaluation method}
We employed various common statistical evaluation metrics, including SRCC, KRCC, PLCC, and RMSE, to assess the performance of our NR-VQA model. These metrics evaluate the monotonicity, linearity, and accuracy of the predictions. In this case, PLCC and RMSE are calculated after a nonlinear four-parameter logistic regression~\cite{seshadrinathan2010study}. All results will be reported on our project GitHub page, while in Table~\ref{tab: ComparisonToSoA} we only report SRCC and PLCC values as below, where $\beta$ parameters are fitted using least squares
\begin{equation}
\label{eq: fx}
    f(x) = \beta_2 + \frac{\beta_1 - \beta_2}{1 + e^{-x + \frac{\beta_3}{|\beta_4|}}} .
\end{equation}

\paragraph{Implementation details} We used ResNet-50~\cite{he2016deep} pre-trained on ImageNet~\cite{krizhevsky2012imagenet} and ViT-B/16~\cite{author2023visualizing} pre-trained models to extract video features. The dataset was split into training and test sets in an 80\%-20\% ratio, respectively. For the LSVQ dataset, we applied 10-fold cross-validation, allocating 9 folds for training and 1 fold for validation, to improve the generalization of our MLP model. The training was conducted over 21 iterations, with 20 epochs for the LSVQ dataset and 120 epochs for other smaller datasets. We used a batch size of 256. Specifically, for the LSVQ dataset, SGD was employed with a learning rate of 0.1 and a weight decay of 0.005, while for other datasets, the learning rate was 0.01 with a weight decay of 0.0005. SWA was implemented in the later stages of training with a learning rate of 0.05. Our loss function utilized weight parameters \( l1_w = 0.6 \) and \( rank_w = 1.0 \).  We tested model selection with two criteria: \(byrmse\), which selects the best model with the smallest RMSE in the validation set, and \( bykrcc\), which selects the best model with the highest KRCC. We evaluated the performance by reporting the median and standard deviation of SRCC and PLCC across 21 iterations. All experiments utilized a workstation with an NVIDIA GeForce RTX 3090 GPU, a 6-core Intel(R) Xeon(R) W-1250 CPU, and 64.0 GB RAM. The proposed model was implemented in PyTorch~\cite{paszke2019pytorch} on Python 3.8.

\subsection{Performance comparison}
\label{ssec: Performance}
We used the following datasets to train and evaluate our proposed model, CVD2014, KoNViD-1k, LIVE-VQC, YouTube-UGC and the largest dataset available, LSVQ. Based on the different strategies for training and testing, we created three versions of our proposed method, ReLaX-VQA:
\begin{enumerate}[label=\roman*, leftmargin=*]
    \item ReLaX-VQA was trained and tested on each dataset (80\% - 20\% random split);
    \item ReLaX-VQA (w/o FT) was trained on LSVQ, and the frozen model was used for testing on the other datasets;
    % \item ReLaX-VQA (w/ FT): this version was trained on the LSVQ dataset and partly on each different dataset so that the saved weights fit the new data features. 
    \item ReLaX-VQA (w/ FT) was trained on LSVQ, and the frozen model was fine-tuned on the other datasets.
\end{enumerate}

% We report results for ReLaX-VQA trained and tested on the same dataset. 
For ReLaX-VQA (w/o FT) and ReLaX-VQA (w/ FT) that were trained on LSVQ, we used the official subset LSVQ\(_{train}\) for training and the LSVQ\(_{test}\) to obtain frozen models. 
% Subsequently, we tested and fine-tuned the trained models on other datasets using the frozen trained models and reported the unfine-tuned test results (\textbf{ReLaX-VQA (w/o FT)}) and fine-tuned test results (\textbf{ReLaX-VQA (w/ FT)}), respectively.
Furthermore, we selected the best-trained model based on \(byrmse\) selection criteria and the results are reported in Table~\ref{tab: ComparisonToSoA}. We compare current state-of-the-art (SOTA) models, including classical statistical and learning-based NR-VQA models, using correlation metrics SRCC, KRCC, PLCC, and RMSE. A key observation is that ReLaX-VQA (w/ FT) demonstrates superior performance across most datasets. Notably, ReLaX-VQA (w/ FT) achieves the highest correlation scores and RMSE values on three out of the four datasets (CVD2014, LIVE-VQC (for PLCC and RMSE), and YouTube-UGC). Furthermore, all versions of ReLaX-VQA significantly outperform traditional NSS-based methods, such as BRISQUE. Even without fine-tuning, ReLaX-VQA (w/o FT) delivers competitive results. For example, on the YouTube-UGC dataset, ReLaX-VQA (w/o FT) records a PLCC of 0.8354 and an RMSE of 0.3768, which are comparable to those of models that have been specifically fine-tuned on the target dataset.  ReLaX-VQA (w/o FT) achieves high scores of SRCC = 0.8312 and PLCC = 0.8427 on KoNViD-1k, not inferior to the best ones that are fine-tuned on the target dataset ($^\dagger$ We need to note that for the VSFA evaluation, we excluded the 110 2160p videos from the YouTube-UGC dataset because our GPU could not support them.). In comparison with the deep learning-based model VSFA, ReLaX-VQA (w/ FT) achieved $\Delta$ SRCC = 0.257 and $\Delta$ PLCC = 0.2982 on LIVE-VQC. Moreover, our model outperforms Fast-VQA-B, which also utilizes fragmentation, on two datasets, CVD2014 and YouTube-UGC. When fine-tuning is applied, ReLaX-VQA (w/ FT) achieves a $\Delta$ PLCC of 0.035 and a $\Delta$ RMSE of -0.0443 on the YouTube-UGC dataset. In comparison to the technical branch of DOVER, ReLaX-VQA (w/ FT) achieved $\Delta\text{SRCC}$ = 0.057 and $\Delta\text{PLCC}$ = 0.0303 on LIVE-VQC. 

It is also worth noting that most of the SOTA methods achieve better performance on single-resolution datasets (i.e., KoNViD-1k), while ReLaX-VQA maintains its high performance on datasets covering multiple resolutions. This indicates that ReLaX-VQA better captures the perception of quality in the vast video parameter space, indicating a wide adaptability and strong generalisation ability.
% Here we report the best results of all selection criteria; see the Appendix for all fine-tuned and unfine-tuned results.

\subsection{Ablation studies, complexity analysis, and limitations}
In our ablation study, we explored the influence of the different components on the performance of our proposed method. In the first part (see Table~\ref{tab: Ablationfragments}), we compare the spatio-temporal features: frame difference and optical flow, along with their performance after residual fragment (RF) sampling and merged fragment (MF). We found that RF sampling of frame difference and optical flow can notably improve performance, especially when using ViT to extract frame difference features, with SRCC and PLCC improving by about 11.92\% and 11.96\%, respectively. The combination of ResNet-50 and ViT enhances performance by about 7.83\% (SRCC) and 7.41\% (PLCC) compared to using ResNet-50 alone, and by about 3.71\% (SRCC) and 3.32\% (PLCC) compared to using ViT alone to extract features. This result emphasizes the effectiveness of multiple DNN integration strategies in dealing with complex video content and the importance of preserving inter-frame spatio-temporal variations for video quality assessment. 

In the second part (Table~\ref{tab: AblationStackVSPool}), we compare the layer stack and pool features of different DNNs. The results show that the layer-stacking framework captures complex video features more effectively compared to the pooling framework. In particular, the combination of ResNet-50 and ViT, ''layer stack + pool features", achieves SRCC of 0.7697 and PLCC of 0.7897, which is noticeably better than using only pooled features. This demonstrates the benefits of the layer-stacking framework in capturing and exploiting deep network features. Here, since ViT is based on the self-attention mechanism, we do not use a layer-stacking framework for this.

% \subsection{Discussion on the Limitations}
% \begin{table*}[t]
%     \centering
%     \caption{FLOPs and running time (on GPU/CPU, average of ten runs) comparison of ReLaX-VQA, and other deep learning SOTA methods across different spatial resolutions.}
%     \label{tab: complexity}
%     \begin{tabular}{@{}l c c c c c c@{}}
%     \toprule
%     \textbf{Method} & \multicolumn{2}{c}{540P} & \multicolumn{2}{c}{720P} & \multicolumn{2}{c}{1080P} \\
%     \cmidrule(lr){2-3} \cmidrule(lr){4-5} \cmidrule(lr){6-7}
%     & FLOPs(G) & Time(s) & FLOPs(G) & Time(s) & FLOPs(G) & Time(s) \\
%     \midrule
%     VSFA~\cite{li2019quality} & 10249  & 4.02/255.59 & 18184 & 6.46/466.88 & 40919 & 13.49/1004.02 \\
%     FAST-VQA-B~\cite{wu2022fast} & 284 & 1.00/24.70 & 284 & 1.18/23.81 & 284 & 1.70/24.74 \\
%     \textbf{ReLaX-VQA (ours)} & 927 & 42.77/64.52 & 927 & 45.61/65.18 & 927 & 54.86/73.84 \\
%     \bottomrule
%     \end{tabular}
% \end{table*}

\begin{table*}[t]
    \centering
    \caption{FLOPs and running time (on CPU, average of ten runs) comparison of ReLaX-VQA and other deep learning SOTA methods across different spatial resolutions.}
    \label{tab: complexity}
    \begin{tabular}{@{}l c c c c c c@{}}
    \toprule
    \textbf{Method} & \multicolumn{2}{c}{540P} & \multicolumn{2}{c}{720P} & \multicolumn{2}{c}{1080P} \\
    \cmidrule(lr){2-3} \cmidrule(lr){4-5} \cmidrule(lr){6-7}
    & FLOPs (G) & Time (s) & FLOPs (G) & Time (s) & FLOPs (G) & Time (s) \\
    \midrule
    VSFA~\cite{li2019quality} & 10249  & 255.59 & 18184 & 466.88 & 40919 & 1004.02 \\
    FAST-VQA-B~\cite{wu2022fast} & 279 & 24.70 & 279 & 23.81 & 279 & 24.74 \\
    DOVER~\cite{wu2023dover} & 282  & 11.95  &292 &12.02 & 282 & 12.48\\
    \textbf{ReLaX-VQA (ours)} & 927 & 31.08 & 927 & 35.93 & 927 & 47.08 \\
    \bottomrule
    \end{tabular}
\end{table*}
% % \hl{Complexity: Runtime or FLOPS comparison?}
% Table~\ref{tab: complexity} compares the three best-performing deep learning-based methods in terms of their computational and runtime complexity. It is evident that ReLaX-VQA and Fast-VQA-B maintain stable FLOPs across resolutions and small differences in CPU and GPU run time.  In contrast, VSFA's FLOPs and running time significantly increase with resolution, with a notably longer running time on the CPU. While FAST-VQA-B has lower FLOPs, ReLaX-VQA achieves a good balance of maintaining a reasonable computational cost while showing consistent efficiency on both the CPU and GPU, demonstrating better generality. It is worth noting that our ReLaX-VQA achieves 57.94 FLOPs per frame in this analysis.

To ensure a fair comparison of the computational efficiency of different methods, all tests were conducted on the same workstation equipped with an NVIDIA GeForce RTX 3090 GPU, a 6-core Intel(R) Xeon(R) W-1250 CPU, and 64 GB of RAM. We used the same video from the KoNViD-1k database for testing at various resolutions. Each test was repeated 10 times, and the average computation time (in seconds) for each method was reported. Table~\ref{tab: complexity} compares the three best-performing deep learning-based methods in terms of their computational and runtime complexity (CPU)~\footnote{Since part of our algorithm (frame differencing, optical flow estimation, and fragment ranking) cannot be fully accelerated on the GPU, including a GPU time comparison would not provide a fair assessment.}. It is evident that ReLaX-VQA, Fast-VQA-B and DOVER maintain stable FLOP numbers across resolutions, resulting in small differences in runtime. For example, ReLaX-VQA requires 31.08 seconds at 540P and 47.08 seconds at 1080P. In contrast, VSFA exhibits a significant increase in both FLOPs and runtime as resolution increases, showing a notably longer runtime on the CPU. While FAST-VQA-B and DOVER are faster, ReLaX-VQA achieves a good balance, maintaining a reasonable computational cost while demonstrating consistent efficiency and better generalization.

One limitation of ReLaX-VQA is the video frame sampling strategy. We sampled two frames every half-second to obtain successive, temporally correlated frame pairs. However, the extent of correlation depends on camera/scene motion, meaning sampled frames have a small chance of not being able to capture the content changes if the camera motion is stationary. Future work will attempt to further optimize this strategy by taking into account the hierarchical encoding of frames within the Group Of Pictures (GOP).
% Another limitation is the large number of parameters, resulting in higher-dimensional feature extraction compared to traditional NR-VQA methods. Notably, our simple MLP design ensures that training and prediction processes do not require significant time, balancing performance and efficiency.
\begin{table}[t]
        \centering
        \caption{Ablation study on \textit{spatio-temporal fragment sampling}.}
        \label{tab: Ablationfragments}
        \begin{tabular}{@{}lccccccccc@{}}
        \toprule
        \textbf{Datasets} & \multicolumn{2}{c}{\textbf{KoNViD-1k}} \\
        \cmidrule(lr){2-3}
        \textbf{Methods/Metrics} & \textbf{SRCC} & \textbf{PLCC}\\
        \midrule
        \(\text{frame difference}\) \textit{(ResNet-50)} & 0.6526 & 0.6525 \\
        \(\text{optical flow}\) \textit{(ResNet-50)} & 0.5793 & 0.5848 \\
        \(\text{frame difference}\) \textit{(ViT)} & 0.6384 & 0.6454 \\
        \(\text{optical flow}\) \textit{(ViT)} & 0.5692 & 0.5844 \\
        \midrule
        \midrule
        \(RF_\text{frame difference}\) \textit{(ResNet-50)} & 0.7082 & 0.7073 \\
        \(RF_{\text{optical flow}}\) \textit{(ResNet-50)} & 0.4844 & 0.4958 \\
        \(RF_{\text{frame difference}}\) \textit{(ViT)} & 0.7145 & 0.7226 \\
        \(RF_{\text{optical flow}}\) \textit{(ViT)} & 0.5032 & 0.5335 \\
        
        \textbf{\textit{MF (ResNet-50)}} & 0.6792 & 0.6788 \\
        \textbf{\textit{MF (ViT)}} & 0.7062 & 0.7057 \\
        \textbf{\textit{MF (ResNet-50+ViT)}} & 0.7324 & 0.7291\\
        \bottomrule
        \end{tabular}
\end{table}
\begin{table}[t]
        \centering
        \caption{Ablation study on \textit{DNN layer stack}.}
        \label{tab: AblationStackVSPool}
        \begin{tabular}{@{}lccccccccc@{}}
        \toprule
        \textbf{Datasets} & \multicolumn{2}{c}{\textbf{KoNViD-1k}} \\
        \cmidrule(lr){2-3}
        \textbf{Metrics/Methods} & \textbf{SRCC} & \textbf{PLCC}\\
        \midrule
        VGG-16 \textit{(\(\text{pool}\))} & 0.6665 & 0.6880 \\
        ResNet-50 \textit{(\(\text{pool}\))} & 0.6751 & 0.7047 \\
        ViT \textit{(\(\text{pool}\))} & 0.6661 & 0.7082 \\
        VGG-16 \textit{(\(\text{layer stack}\))} & 0.7513 & 0.7700 \\
        ResNet-50 \textit{(\(\text{layer stack}\))} & 0.7630 & 0.7828 \\
        \midrule
        \midrule
        VGG-16+ViT \textit{(\(\text{pool+pool}\))} & 0.7100 & 0.7364 \\
        ResNet-50+ViT \textit{(\(\text{pool+pool}\))} & 0.7253 & 0.7469 \\
        VGG-16+ViT \textit{(\(\text{layer stack+pool}\))} & 0.7650 & 0.7827 \\
        ResNet-50+ViT \textit{(\(\text{layer stack+pool}\))} & 0.7697 & 0.7897 \\
        \bottomrule
        \end{tabular} 
\end{table}

\section{Conclusion and future work}
\label{sec: Conlcusions}
In this paper, we propose ReLaX-VQA to address the challenges of NR-VQA for UGC videos. We employed residual difference and optical flow fragments between successive sampled frames along with a ranking process to capture the spatio-temporal information. Additionally, we enhanced DNN features by stacking layers to capture local and global information affecting video quality. We trained and tested our model on four public UGC datasets. Our results showed that ReLaX-VQA outperforms existing NR-VQA methods and demonstrates consistently high performance across the different datasets, achieving an overall average SRCC value of 0.8658 and PLCC value of 0.8873. Our method verified the benefit of residual information in enhancing the perceptual capability of NR-VQA models. Future research will focus on improving efficiency through network architectures (e.g., Swin Transformer or Mamba for feature extraction) and enhancing regression approaches (e.g., KAN) to optimise memory usage and processing speed.
% Our results showed that ReLaX-VQA outperforms existing NR-VQA methods and demonstrates consistently high performance across the different datasets, 
% or implementing end-to-end training to improve performance and speed trade-offs. 
% Further directions include adding semantic features and exploring multimodal models for VQA.

\section*{Acknowledgment}
This work was funded by the UKRI MyWorld Strength in Places Programme (SIPF00006/1).

\bibliographystyle{IEEEtran}
\bibliography{ref}

\begin{thebibliography}{10}
\providecommand{\url}[1]{#1}
\csname url@samestyle\endcsname
\providecommand{\newblock}{\relax}
\providecommand{\bibinfo}[2]{#2}
\providecommand{\BIBentrySTDinterwordspacing}{\spaceskip=0pt\relax}
\providecommand{\BIBentryALTinterwordstretchfactor}{4}
\providecommand{\BIBentryALTinterwordspacing}{\spaceskip=\fontdimen2\font plus
\BIBentryALTinterwordstretchfactor\fontdimen3\font minus \fontdimen4\font\relax}
\providecommand{\BIBforeignlanguage}[2]{{%
\expandafter\ifx\csname l@#1\endcsname\relax
\typeout{** WARNING: IEEEtran.bst: No hyphenation pattern has been}%
\typeout{** loaded for the language `#1'. Using the pattern for}%
\typeout{** the default language instead.}%
\else
\language=\csname l@#1\endcsname
\fi
#2}}
\providecommand{\BIBdecl}{\relax}
\BIBdecl

\bibitem{sandvine}
Sanvdine, ``The global internet phenomena report,'' \url{https://www.applogicnetworks.com/phenomena?submissionGuid=bd3de666-249c-4927-9bab-b84c3577b2c9}, [Online; accessed 2024].

\bibitem{Adsumilli2019}
\BIBentryALTinterwordspacing
B.~Adsumilli, S.~Inguva, Y.~Wang, J.~Huoponen, and R.~Wolf, ``{Launching a YouTube dataset of user-generated content},'' 2019. [Online]. Available: \url{https://www.googblogs.com/launching-a-youtube-dataset-of-user-generated-content/}
\BIBentrySTDinterwordspacing

\bibitem{wang2019youtube}
Y.~Wang, S.~Inguva, and B.~Adsumilli, ``Youtube ugc dataset for video compression research,'' in \emph{IEEE MMSP}, 2019.

\bibitem{wang2004image}
Z.~Wang, A.~C. Bovik, H.~R. Sheikh, and E.~P. Simoncelli, ``Image quality assessment: from error visibility to structural similarity,'' \emph{IEEE TIP}, vol.~13, no.~4, pp. 600--612, 2004.

\bibitem{li2016toward}
Z.~Li, A.~Aaron, I.~Katsavounidis, A.~Moorthy, and M.~Manohara, ``Toward a practical perceptual video quality metric,'' \emph{The Netflix Tech Blog}, vol.~6, no.~2, 2016.

\bibitem{alzubaidi2021review}
L.~Alzubaidi, J.~Zhang, A.~J. Humaidi, A.~Al-Dujaili, Y.~Duan, O.~Al-Shamma, J.~Santamar{\'\i}a, M.~A. Fadhel, M.~Al-Amidie, and L.~Farhan, ``Review of deep learning: concepts, cnn architectures, challenges, applications, future directions,'' \emph{Journal of big Data}, vol.~8, pp. 1--74, 2021.

\bibitem{li2019quality}
D.~Li, T.~Jiang, and M.~Jiang, ``Quality assessment of in-the-wild videos,'' in \emph{ACM MM}, 2019, pp. 2351--2359.

\bibitem{tu2021rapique}
Z.~Tu, X.~Yu, Y.~Wang, N.~Birkbeck, B.~Adsumilli, and A.~C. Bovik, ``Rapique: Rapid and accurate video quality prediction of user generated content,'' \emph{IEEE OJSP}, vol.~2, pp. 425--440, 2021.

\bibitem{ying2021patch}
Z.~Ying, M.~Mandal, D.~Ghadiyaram, and A.~Bovik, ``Patch-vq:'patching up'the video quality problem,'' in \emph{IEEE/CVF CVPR}, 2021, pp. 14\,019--14\,029.

\bibitem{liu2018end}
W.~Liu, Z.~Duanmu, and Z.~Wang, ``End-to-end blind quality assessment of compressed videos using deep neural networks.'' in \emph{ACM Multimedia}, 2018, pp. 546--554.

\bibitem{sun2022deep}
W.~Sun, X.~Min, W.~Lu, and G.~Zhai, ``A deep learning based no-reference quality assessment model for ugc videos,'' in \emph{ACM MM}, 2022, pp. 856--865.

\bibitem{wu2023discovqa}
H.~Wu, C.~Chen, L.~Liao, J.~Hou, W.~Sun, Q.~Yan, and W.~Lin, ``Discovqa: Temporal distortion-content transformers for video quality assessment,'' \emph{IEEE TCSVT}, 2023.

\bibitem{zhao2023zoom}
K.~Zhao, K.~Yuan, M.~Sun, and X.~Wen, ``Zoom-vqa: Patches, frames and clips integration for video quality assessment,'' in \emph{IEEE/CVF CVPR}, 2023, pp. 1302--1310.

\bibitem{wu2022fast}
H.~Wu, C.~Chen, J.~Hou, L.~Liao, A.~Wang, W.~Sun, Q.~Yan, and W.~Lin, ``Fast-vqa: Efficient end-to-end video quality assessment with fragment sampling,'' in \emph{ECCV}, 2022, pp. 538--554.

\bibitem{sze2017efficient}
V.~Sze, Y.-H. Chen, T.-J. Yang, and J.~S. Emer, ``Efficient processing of deep neural networks: A tutorial and survey,'' \emph{Proceedings of the IEEE}, vol. 105, no.~12, pp. 2295--2329, 2017.

\bibitem{you2021long}
J.~You, ``Long short-term convolutional transformer for no-reference video quality assessment,'' in \emph{ACM MM}, 2021, pp. 2112--2120.

\bibitem{madhusudana2023conviqt}
P.~C. Madhusudana, N.~Birkbeck, Y.~Wang, B.~Adsumilli, and A.~C. Bovik, ``Conviqt: Contrastive video quality estimator,'' \emph{IEEE TIP}, 2023.

\bibitem{r:h264}
{ITU-T Rec. H.264}, ``Advanced video coding for generic audiovisual services,'' ITU-T, 2005.

\bibitem{r:HEVC}
{ITU-T Rec H.265}, ``High efficiency video coding,'' ITU-T, 2015.

\bibitem{he2016deep}
K.~He, X.~Zhang, S.~Ren, and J.~Sun, ``Deep residual learning for image recognition,'' in \emph{Proceedings of the IEEE CVPR}, 2016, pp. 770--778.

\bibitem{vaswani2017attention}
A.~Vaswani, N.~Shazeer, N.~Parmar, J.~Uszkoreit, L.~Jones, A.~N. Gomez, {\L}.~Kaiser, and I.~Polosukhin, ``Attention is all you need,'' \emph{Advances in NEURIPS}, vol.~30, 2017.

\bibitem{nuutinen2016cvd2014}
M.~Nuutinen, T.~Virtanen, M.~Vaahteranoksa, T.~Vuori, P.~Oittinen, and J.~H{\"a}kkinen, ``{CVD2014—A} database for evaluating no-reference video quality assessment algorithms,'' \emph{IEEE TIP}, vol.~25, no.~7, pp. 3073--3086, 2016.

\bibitem{hosu2017konstanz}
V.~Hosu, F.~Hahn, M.~Jenadeleh, H.~Lin, H.~Men, T.~Szir{\'a}nyi, S.~Li, and D.~Saupe, ``{The Konstanz natural video database (KoNViD-1k)},'' in \emph{International Conference on Quality of Multimedia Experience}, 2017.

\bibitem{sinno2018large}
Z.~Sinno and A.~C. Bovik, ``Large-scale study of perceptual video quality,'' \emph{IEEE TIP}, vol.~28, no.~2, pp. 612--627, 2018.

\bibitem{dodge2016understanding}
S.~Dodge and L.~Karam, ``Understanding how image quality affects deep neural networks,'' in \emph{2016 QoMEX}.

\bibitem{ghadiyaram2017capture}
D.~Ghadiyaram, J.~Pan, A.~C. Bovik, A.~K. Moorthy, P.~Panda, and K.-C. Yang, ``In-capture mobile video distortions: A study of subjective behavior and objective algorithms,'' \emph{IEEE TCSVT}, vol.~28, no.~9, pp. 2061--2077, 2017.

\bibitem{gotz2019no}
F.~G{\"o}tz-Hahn, V.~Hosu, H.~Lin, and D.~Saupe, ``No-reference video quality assessment using multi-level spatially pooled features,'' \emph{arXiv preprint arXiv:1912.07966}, 2019.

\bibitem{mittal2012making}
A.~Mittal, R.~Soundararajan, and A.~C. Bovik, ``Making a “completely blind” image quality analyzer,'' \emph{IEEE Signal Processing Letters}, vol.~20, no.~3, pp. 209--212, 2012.

\bibitem{moorthy2011blind}
A.~K. Moorthy and A.~C. Bovik, ``Blind image quality assessment: From natural scene statistics to perceptual quality,'' \emph{IEEE TIP}, vol.~20, no.~12, pp. 3350--3364, 2011.

\bibitem{mittal2011blind}
A.~Mittal, A.~K. Moorthy, and A.~C. Bovik, ``Blind/referenceless image spatial quality evaluator,'' in \emph{ASILOMAR}, 2011, pp. 723--727.

\bibitem{6705673}
M.~A. Saad, A.~C. Bovik, and C.~Charrier, ``Blind prediction of natural video quality,'' \emph{IEEE TIP}, vol.~23, no.~3, pp. 1352--1365, 2014.

\bibitem{7025098}
J.~Xu, P.~Ye, Y.~Liu, and D.~Doermann, ``No-reference video quality assessment via feature learning,'' in \emph{2014 IEEE International Conference on Image Processing (ICIP)}, 2014, pp. 491--495.

\bibitem{korhonen2019two}
J.~Korhonen, ``Two-level approach for no-reference consumer video quality assessment,'' \emph{IEEE TIP}, vol.~28, no.~12, pp. 5923--5938, 2019.

\bibitem{tu2021ugc}
Z.~Tu, Y.~Wang, N.~Birkbeck, B.~Adsumilli, and A.~C. Bovik, ``Ugc-vqa: Benchmarking blind video quality assessment for user generated content,'' \emph{IEEE TIP}, vol.~30, pp. 4449--4464, 2021.

\bibitem{li2021unified}
D.~Li, T.~Jiang, and M.~Jiang, ``Unified quality assessment of in-the-wild videos with mixed datasets training,'' \emph{International Journal of Computer Vision}, vol. 129, no.~4, pp. 1238--1257, 2021.

\bibitem{liu2024scaling}
Y.~Liu, Y.~Quan, G.~Xiao, A.~Li, and J.~Wu, ``Scaling and masking: A new paradigm of data sampling for image and video quality assessment,'' \emph{arXiv preprint arXiv:2401.02614}, 2024.

\bibitem{ke2023mret}
J.~Ke, T.~Zhang, Y.~Wang, P.~Milanfar, and F.~Yang, ``Mret: Multi-resolution transformer for video quality assessment,'' \emph{Frontiers in Signal Processing}, vol.~3, p. 1137006, 2023.

\bibitem{lao2022attentions}
S.~Lao, Y.~Gong, S.~Shi, S.~Yang, T.~Wu, J.~Wang, W.~Xia, and Y.~Yang, ``Attentions help cnns see better: Attention-based hybrid image quality assessment network,'' in \emph{IEEE/CVF CVPR}, 2022, pp. 1140--1149.

\bibitem{zhang2022hvs}
A.-X. Zhang, Y.-G. Wang, W.~Tang, L.~Li, and S.~Kwong, ``Hvs revisited: A comprehensive video quality assessment framework,'' \emph{arXiv preprint arXiv:2210.04158}, 2022.

\bibitem{Feng_2024_WACV}
C.~Feng, D.~Danier, F.~Zhang, and D.~Bull, ``Rankdvqa: Deep vqa based on ranking-inspired hybrid training,'' in \emph{Proceedings of the IEEE/CVF Winter Conference on Applications of Computer Vision (WACV)}, January 2024, pp. 1648--1658.

\bibitem{li2022blindly}
B.~Li, W.~Zhang, M.~Tian, G.~Zhai, and X.~Wang, ``Blindly assess quality of in-the-wild videos via quality-aware pre-training and motion perception,'' \emph{IEEE TCSVT}, vol.~32, no.~9, pp. 5944--5958, 2022.

\bibitem{wang2021rich}
Y.~Wang, J.~Ke, H.~Talebi, J.~G. Yim, N.~Birkbeck, B.~Adsumilli, P.~Milanfar, and F.~Yang, ``Rich features for perceptual quality assessment of ugc videos,'' in \emph{CVPR}, 2021, pp. 13\,435--13\,444.

\bibitem{wu2023neighbourhood}
H.~Wu, C.~Chen, L.~Liao, J.~Hou, W.~Sun, Q.~Yan, J.~Gu, and W.~Lin, ``Neighbourhood representative sampling for efficient end-to-end video quality assessment,'' \emph{IEEE Transactions on Pattern Analysis and Machine Intelligence}, 2023.

\bibitem{wu2023dover}
H.~Wu, E.~Zhang, L.~Liao, C.~Chen, J.~H. Hou, A.~Wang, W.~S. Sun, Q.~Yan, and W.~Lin, ``Exploring video quality assessment on user generated contents from aesthetic and technical perspectives,'' in \emph{International Conference on Computer Vision (ICCV)}, 2023.

\bibitem{farneback2003two}
G.~Farneb{\"a}ck, ``Two-frame motion estimation based on polynomial expansion,'' in \emph{Image Analysis: 13th Scandinavian Conference}, 2003, pp. 363--370.

\bibitem{wang2023ugc}
X.~Wang, A.~Katsenou, and D.~Bull, ``Ugc quality assessment: exploring the impact of saliency in deep feature-based quality assessment,'' in \emph{Applications of Digital Image Processing XLVI}, vol. 12674.\hskip 1em plus 0.5em minus 0.4em\relax SPIE, 2023, pp. 351--365.

\bibitem{ramsookicip2023}
D.~Ramsook and A.~Kokaram, ``Learnt deep hyperparameter selection in adversarial training for compressed video enhancement with a perceptual critic,'' in \emph{IEEE ICIP}, 2023, pp. 2420--2424.

\bibitem{rosenblatt1958perceptron}
F.~Rosenblatt, ``The perceptron: a probabilistic model for information storage and organization in the brain.'' \emph{Psychological review}, vol.~65, no.~6, p. 386, 1958.

\bibitem{loshchilov2016sgdr}
I.~Loshchilov and F.~Hutter, ``Sgdr: Stochastic gradient descent with warm restarts,'' \emph{arXiv preprint arXiv:1608.03983}, 2016.

\bibitem{izmailov2018averaging}
P.~Izmailov, D.~Podoprikhin, T.~Garipov, D.~Vetrov, and A.~G. Wilson, ``Averaging weights leads to wider optima and better generalization,'' \emph{arXiv preprint arXiv:1803.05407}, 2018.

\bibitem{wen2021strong}
S.~Wen and J.~Wang, ``A strong baseline for image and video quality assessment,'' \emph{arXiv preprint arXiv:2111.07104}, 2021.

\bibitem{seshadrinathan2010study}
K.~Seshadrinathan, R.~Soundararajan, A.~C. Bovik, and L.~K. Cormack, ``Study of subjective and objective quality assessment of video,'' \emph{IEEE TIP}, vol.~19, no.~6, pp. 1427--1441, 2010.

\bibitem{krizhevsky2012imagenet}
A.~Krizhevsky, I.~Sutskever, and G.~E. Hinton, ``Imagenet classification with deep convolutional neural networks,'' \emph{Advances in NEURIPS}, vol.~25, 2012.

\bibitem{author2023visualizing}
\BIBentryALTinterwordspacing
A.~Jadon, ``Visualizing attention in vision transformer,'' \emph{Medium}, Jan 2023. [Online]. Available: \url{https://ai.plainenglish.io/visualizing-attention-in-vision-transformer-c871908d86de}
\BIBentrySTDinterwordspacing

\bibitem{paszke2019pytorch}
A.~Paszke, S.~Gross, F.~Massa, A.~Lerer, J.~Bradbury, G.~Chanan, T.~Killeen, Z.~Lin, N.~Gimelshein, L.~Antiga \emph{et~al.}, ``Pytorch: An imperative style, high-performance deep learning library,'' \emph{Advances in NEURIPS}, vol.~32, 2019.

\end{thebibliography}

\end{document}